\documentclass[iop,apj]{emulateapj}
\usepackage{graphicx,multirow,amsmath}
\usepackage{color}
\usepackage{natbib}
\usepackage{url}
\usepackage{hyperref}
\usepackage{breakurl}
\slugcomment{Accepted to ApJ; 2015 June 8}
\shorttitle{Radio and X-ray Observations of the Galactic Center Magnetar}
\shortauthors{T.~T.~Pennucci~et.~al.}

\def\gcm {J1745$-$2900}
\def\sgrgcm {SGR~1745$-$2900}
\def\xmm {\emph{XMM-Newton}}
\def\cxo {\emph{Chandra}}
\def\swift {\emph{Swift}}

\def\flux {\mbox{erg cm$^{-2}$ s$^{-1}$}}
\def\lum {\mbox{erg s$^{-1}$}}
\def\nh {$N_{\rm H}$}
\def\dmunit {cm$^{-3}$~pc}

\begin{document}

\title{Simultaneous Multi-band Radio \& X-ray Observations of\\the Galactic Center Magnetar SGR~1745$-$2900}

\author{
    T.~T.~Pennucci\altaffilmark{1},
    A.~Possenti\altaffilmark{2},
    P.~Esposito\altaffilmark{3,4},
    N.~Rea\altaffilmark{5,6},
    D.~Haggard\altaffilmark{7},\\
    F.~K.~Baganoff\altaffilmark{8},
    M.~Burgay\altaffilmark{2},
    F.~Coti~Zelati\altaffilmark{9,5,10},
    G.~L.~Israel\altaffilmark{11},
    A.~Minter\altaffilmark{12}}

\altaffiltext{1}{University of Virginia, Department of Astronomy, PO Box 400325 Charlottesville, VA 22904-4325, USA, \href{mailto:pennucci@virginia.edu}{\tt pennucci@virginia.edu}}
\altaffiltext{2}{Osservatorio Astronomico di Cagliari, via della Scienza 5, 09047, Cagliari, Italy}
\altaffiltext{3}{Istituto di Astrofisica Spaziale e Fisica Cosmica - Milano, INAF, via E. Bassini 15, I-20133, Milano, Italy}
\altaffiltext{4}{Harvard--Smithsonian Center for Astrophysics, 60 Garden Street, Cambridge, MA 02138, USA}
\altaffiltext{5}{Anton Pannekoek Institute for Astronomy, University of Amsterdam, Postbus 94249, NL-1090-GE Amsterdam, The Netherlands}
\altaffiltext{6}{Institute of Space Sciences (ICE, CSIC-IEEC), Carrer de Can Magrans, S/N, 08193, Barcelona, Spain}
\altaffiltext{7}{Department of Physics and Astronomy, Amherst College, Amherst, MA 01002-5000, USA}
\altaffiltext{8}{Kavli Institute for Astrophysics and Space Research, Massachusetts Institute of Technology, Cambridge, MA 02139, USA}
\altaffiltext{9}{Universit\`a dell'Insubria, via Valleggio 11, I-22100 Como, Italy}
\altaffiltext{10}{INAF--Osservatorio Astronomico di Brera, via Bianchi 46, I-23807, Merate (LC), Italy}
\altaffiltext{11}{INAF--Osservatorio Astronomico di Roma, via Frascati 33, I-00040, Monteporzio Catone, Roma, Italy}
\altaffiltext{12}{National Radio Astronomy Observatory, PO Box 2, 155 Observatory Road, Green Bank, WV 24944, USA}

\begin{abstract}
We report on multi-frequency, wideband radio observations of the Galactic Center magnetar (SGR~1745$-$2900) with the Green Bank Telescope for $\sim$100~days immediately following its initial X-ray outburst in April 2013.  We made multiple simultaneous observations at 1.5, 2.0, and 8.9~GHz, allowing us to examine the magnetar's flux evolution, radio spectrum, and interstellar medium parameters (such as the dispersion measure (DM), the scattering timescale and its index).  During two epochs, we have simultaneous observations from the \cxo\ X-ray Observatory, which permitted the absolute alignment of the radio and X-ray profiles.  As with the two other radio magnetars with published alignments, the radio profile lies within the broad peak of the X-ray profile, preceding the X-ray profile maximum by $\sim$0.2 rotations.  We also find that the radio spectral index $\gamma$ is significantly negative between $\sim$2 and 9~GHz; during the final $\sim$30~days of our observations $\gamma \sim -1.4$, which is typical of canonical pulsars.  The radio flux has not decreased during this outburst, whereas the long-term trends in the other radio magnetars show concomitant fading of the radio and X-ray fluxes.  Finally, our wideband measurements of the DMs taken in adjacent frequency bands in tandem are stochastically inconsistent with one another.  Based on recent theoretical predictions, we consider the possibility that the dispersion measure is frequency-dependent.  Despite having several properties in common with the other radio magnetars, such as $L_{\textrm{X,qui}}/L_{\textrm{rot}} \lesssim 1$, an increase in the radio flux during the X-ray flux decay has not been observed thus far in other systems.
\end{abstract}

\keywords{Galaxy: center --- pulsars: individual (PSR~\gcm, SGR~1745$-$2900) --- stars: magnetars} 

\section{Introduction}
\label{intro}

Magnetars are exotica among the exotic: whereas other pulsars are sustained by their stored angular momentum, the primary energy source that powers this special class of objects is likely the neutron star's immense magnetic field \citep{Mereghetti15}.  The field strengths take on the highest values ever inferred, typically $> 10^{12}$~G and even up to $\sim 10^{15}$~G.  According to the McGill Online Magnetar Catalog\footnote{\url{http://www.physics.mcgill.ca/~pulsar/magnetar/main.html}} \citep{Olausen14}, there are 28 known magnetars, of which only four have displayed pulsed radio emission.

\sgrgcm\ (J1745$-$2900, hereafter) is the most recent addition to the small collection of magnetars with observed pulsed radio emission (the ``radio magnetars'', to which we will refer by their PSR names: J1809$-$1943 (XTE~1810$-$197), J1550$-$5418 (1E~1547.0$-$5408), \& J1622$-$4950 \citep{Camilo06,Camilo07c,Levin10}).  On 25 April 2013, one day after the XRT aboard the \swift\ satellite detected flaring activity coincident with the Galactic Center \citep{Degenaar13}, a short X-ray burst was observed by \swift/BAT showing characteristics similar to those usually observed from soft gamma-ray repeaters \citep{Kennea13a}.  Shortly thereafter, observations from the \textit{NuSTAR} satellite identified the source as a magnetar with a $P_s = 3.76$~s spin period, and its radio pulsations were subsequently seen by the Effelsberg 100-m Telescope \citep{Mori13a,Mori13b,Eatough13a}.  \gcm\ was soon physically associated with the Galactic Center, located only $\sim$2.5'' away from Sagittarius~A* (Sgr~A*) with a neutral hydrogen column density and dispersion measure (DM) consistent with being within $\sim$2~pc of the Milky Way's central black hole \citep{Eatough13b,Rea13}.

Early determinations of its spin-down $\dot{P}_s$ put \gcm\ squarely within the magnetar population, having an inferred magnetic field strength at the equator $B_s \sim 3.2 \times 10^{19}$~G~$\sqrt{P_s\dot{P}_s} \sim 1.6 \times 10^{14}$~G, a characteristic age $\tau_c \sim P_s/(2\dot{P}_s) \sim 9$~kyr, and a spin-down luminosity of $\dot{E} = L_{\textrm{rot}} = 3.95 \times 10^{46}$~erg~s$^{-1} (P_s^{-3}\dot{P}_s) \sim 4.9 \times 10^{33}$~erg~s$^{-1}$ \citep{Rea13}.  However, its estimated quiescent X-ray luminosity of $L_{\textrm{X,qui}} < 10^{34}$~erg~s$^{-1}$ \citep{Zelati15} may place \gcm\ on the side of $L_{\textrm{X,qui}}/L_{\textrm{rot}} < 1$, opposite the ``classic magnetars'' but alongside the other three radio magnetars, high-$B$ pulsars, and radio pulsars with X-ray emission \citep{Rea12a}.

Given the unique environment in which \gcm\ resides, the detection of its radio pulses is somewhat surprising.   Indeed, numerous surveys of the Galactic Center region covering $\sim$1--20~GHz have failed to find a pulsar within the central parsec \citep[most recently,][]{Johnston06,Deneva09,Macquart10,Bates11,Siemion13}.  The discovery of this single magnetar has led to a windfall of implications for future discoveries \citep{Chennamangalam14,Dexter14,Macquart15}.  Because of its proximity to the Galactic Center, \gcm\ has the largest DM (1778~\dmunit) and rotation measure ($-6.696 \times 10^4$~rad~m$^{-2}$) of any known pulsar \citep{Eatough13b}.  The predicted value for the scattering timescale at 1~GHz, based on empirical relationships given its DM, is $\sim$ 1000~s \citep{Krishnakumar15,Lewandowski15a}, meaning that \gcm\ would be undetectable at frequencies less than $\sim$5~GHz.  The situation is exacerbated by the presence of an additional scattering screen in the Galactic Center \citep{CordesLazio97}.  Normally, the prospect of detecting distant radio pulsars above several GHz is bleak, since their average spectral index is $\sim$ -1.4 \citep{Bates13}.  However, because the other radio magnetars have flat/inverted spectra, one might expect to detect \gcm's unscattered pulse profile at high frequencies.  In the analyses that follow, we will reiterate the finding that \gcm\ has a significantly smaller scattering timescale than predicted \citep{Spitler14}, and will show that \gcm\ was much brighter at lower frequencies, having a very negative spectral index some 100~days after the onset of its outburst, even though more recent observations by \citet{Torne15} showed the spectral index has since flattened.

In this paper, we analyze multi-frequency radio data over the first $\sim$100~days after \gcm's discovery, during which time there were two additional \swift/BAT-detected bursts on 7 June 2013 and 5 August 2013 \citep{Kennea13b,Kennea13c}.  For two of our epochs, which bracket the third burst by $\sim$1~week on either side, we have simultaneous \cxo\ observations.  These observations allow us to find the absolute alignment of the radio and X-ray profiles, and to look for correlated events.  We comment on the spin evolution and timing, and examine the profile stability, the radio flux evolution, and the radio spectrum.  Finally, we make global models of the profile evolution across the low frequency bands in order to examine the temporal and frequency dependencies of the scattering timescale and dispersion measure.  We then discuss characteristics of this source in comparison with other radio-loud magnetars.
\\\\\\

\section{Observations}
\label{obs}

\subsection{Radio}
\label{radio_obs}

We made early detections of \gcm\ during fourteen observing epochs with the 100-m Robert C. Byrd Green Bank Telescope (GBT) in three different frequency bands with various overlap: 1.1--1.9~GHz (5~epochs), 1.6--2.4~GHz (7~epochs), and 8.5--9.3~GHz (11~epochs) (PI: A. Possenti).  Because each observation covers a large bandwidth, we refer to each set of data based on the IEEE radio band for which each of the receiver systems is named (``L-band'', ``S-band'', or ``X-band'', respectively), instead of referring to specific (central) frequencies.  Table~\ref{obs_table} contains details of the observations.  In all cases, we observed using the Green Bank Ultimate Pulsar Processing Instrument \citep[GUPPI\footnote{\url{www.safe.nrao.edu/wiki/bin/view/CICADA/NGNPP}},][]{Duplain08} in ``incoherent search mode'', recording dual-polarization time-series data in 2048 frequency channels with a temporal resolution of 0.65536~ms.

Each epoch's data were folded with the pulsar software library \textsc{dspsr}\footnote{\url{http://dspsr.sourceforge.net/}} using a nominal ephemeris with a constant spin frequency (see \S\ref{timing}) and the \cxo-determined position $\rm \alpha_{J2000.0}=17^h45^m40\fs169,~\delta_{J2000.0}~=~29\degr00'29\farcs84$ \citep{Rea13}.  The data were initially folded into 1 min subintegrations, with 2048~profile phase bins across 128 frequency channels.  We adopted the published dispersion measure value of 1778~\dmunit\ for averaging frequency channels together \citep{Eatough13b}.  Persistent, narrow-band radio frequency interference (RFI) was excised automatically; any remaining significantly corrupted channels or subintegrations were removed from the data by hand.

Calibration scans were taken for each observation using the local noise diode, pulsed at 25~Hz while on source.  We recorded on- and off-source scans of a standard flux calibrator (QSO~B1442+101) in each frequency band only during the final epoch (MJD 56516).  We have used this one set of flux calibration scans to calibrate the whole data set.  Standard programs from the \textsc{PSRCHIVE}\footnote{\url{http://psrchive.sourceforge.net/}} pulsar software library \citep{Hotan04,vStrat12} were used to calibrate the absolute flux density scale of the noise diode, which is then used to determine the magnetar's flux density\footnote{The \textsc{PSRCHIVE} calibration process produced unphysical results for the earliest S-band detection (MJD 56424); we have calibrated it by using an approximation based on the measured S-band system equivalent flux density and the radiometer equation (cf. \S7.3.2 of \citet{L&K04}).  The result is reasonable, given that the next S-band observation five days later has a comparable flux density (see Figure~\ref{flux_evol}).}.

\begin{figure}
\epsscale{1.0}
\plotone{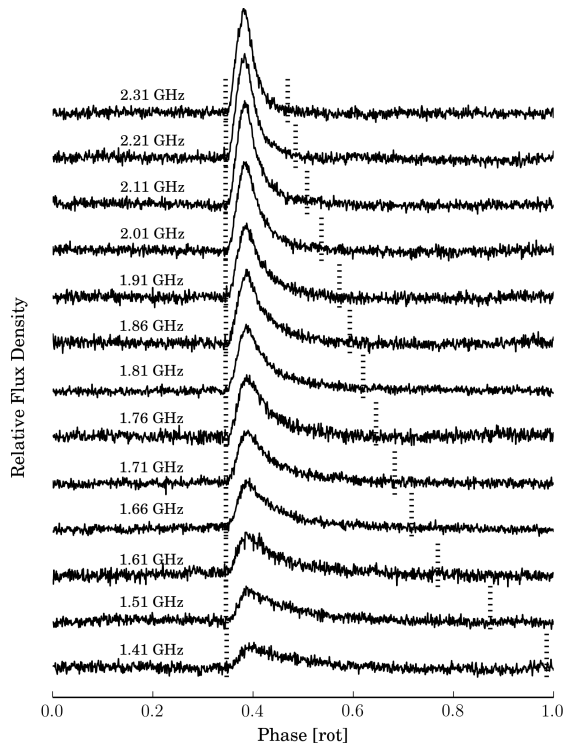}
\caption{Examples of L- and S- band profiles averaged over all epochs.  The profiles are shown with 1024~phase~bins for clarity.  These data are aligned via a wideband portrait model, as described in \S\ref{wideband}.  In general, the un-averaged profiles were also of good quality, with only minor systematics in the baseline.  The total bandwidth covered across these two bands is about 1~GHz, from $\sim$1.4 to 2.4 GHz; 25~MHz of data were averaged for each of these profiles, with their center frequencies shown.  The profiles were very well described by a single scattered Gaussian component, and so we do not over-plot the wideband model.  The vertical dotted lines show examples of on-pulse regions used for the flux density measurements.  See \S\ref{flux} for details.}
\label{LSprofs}
\end{figure}

The combination of the large amount of observed scattering (\S\ref{wideband_stuff}), the pulsar's spectrum (\S\ref{spectrum}, Figure~\ref{spectrum_56516}), receiver roll-off, and the presence of gain variations (see below) rendered significant portions of the ends of L-band useless.  Namely, there was no pulsed signal in the lower 300~MHz portion of L-band, which we masked from further analysis, along with the top 50~MHz (which is part of the overlap with S-band).  In combination with the narrow-band RFI, this left less than $\sim$400~MHz of clean, usable bandwidth.  Similarly, at S-band we had to remove the lower $\sim$100~MHz and the upper $\sim$25~MHz, and in total $\sim$625~MHz of usable band remained\footnote{In two epochs, however, instrument problems left only half of S-band viable.  See Table~\ref{obs_table}.}.  Only 3\% of the data was clipped from either end of X-band, with a total of 10\% removed.  We took these seemingly draconian measures to offset the original data quality and to ensure that the time- and frequency-averaged profiles were of reasonably high quality (e.g., see Figure~\ref{LSprofs}).  This was enabled by the source's relatively large flux density. 

\begin{figure}
\epsscale{1.0}
\plotone{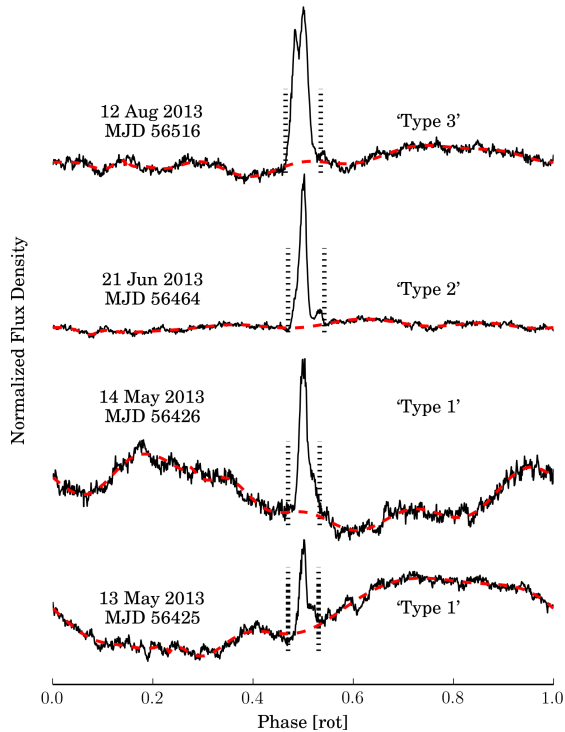}
\caption{Examples of time- and frequency-averaged X-band profiles.  The profiles are shown with 1024~phase~bins for clarity.  The baseline variations were removed on a profile-to-profile basis by fitting a high degree polynomial (red dashed lines) to the off-pulse region (outside the dotted lines) in order to make measurements of the flux density (see \S\ref{flux}).  The on-pulse phase window varied in size between about 6 and 8\%.  The profile evolved monotonically from one ``type'' to the next (see \S\ref{prof_var} and Table~\ref{obs_table}).}
\label{Xprofs}
\end{figure}

The data quality situation at X-band was still more complicated.  As also noted by \citet{Lynch14} in their investigation of this magnetar, large gain variations on timescales from a fraction of a pulse period to several seconds (visible in the time-series data) are prevalent in X-band at the GBT, when pointed at the Galactic Center.  The variations did not (necessarily) integrate away over hour-long observations and are representative of a stochastic red-noise process.  We attribute these variations to changes in atmospheric opacity \citep{Lynch14} and/or small pointing errors, noting a strong resonance in the GBT X-band pointing very near 0.3~Hz\footnote{Even though the average pointing errors at X-band are only on the order of several arcseconds at mid-elevations and mild wind conditions, the power spectrum in elevation offset shows resonances overlapping with the magnetar's spin frequency ($0.27...$~Hz).  See \url{http://www.gb.nrao.edu/~rmaddale/GBT/Commissioning/Pointing_Gregorian_HighFreq/PntStabilityXBand.pdf} for details.}.  The gain variations would be manifested by the relatively small beam of X-band ($\sim$1.4$'$, compared to $\sim$6$'$ and $\sim$9$'$ for S- and L-band) oscillating over the crowded, bright Galactic Center (the central parsec extends $\sim$0.4$'$, and the separation of \gcm\ from Sgr~A* is only $\sim$0.04$'$ \citep{Rea13}).  Additionally, it is likely that the baseline variations are much less prominent at low frequencies because they act as ``zero-DM'' signals that get smeared out when the pulsar's signal is dedispersed.  \citet{Lynch14} also state that the effect may be a function of elevation, which fits with our pointing-resonance hypothesis, since the influence of variable elements like the wind will be a function of elevation.  The persistence and variability of these variations can be seen in Figure~\ref{Xprofs}.

The analyses that follow utilized these folded profiles in a variety of reduced forms.  Unless otherwise noted, the reduced radio data have 2048 profile bins ($\sim$7.2~ms per bin), 32 frequency channels (25~MHz per channel), and 5~min subintegrations; in this work, we only consider the total intensity profiles.

\setlength{\tabcolsep}{5pt}
\tabletypesize{\scriptsize}
\begin{deluxetable}{cccc}
    \centering
    \tablewidth{0pt}
    \tablecaption{\label{obs_table}Summary of GBT Observations} 
    \tablehead{\colhead{UTC}  &  \colhead{MJD}   &   \colhead{Bands}    &   \colhead{Approx. Length} \\
               \colhead{Epoch}  &  \colhead{} &   \colhead{Observed} &   \colhead{[min]}}
    \startdata
    2013-05-04  &   56416   &   X           &   20 \\
    2013-05-12  &   56424   &   S*,X        &   122,200 \\
    2013-05-13  &   56425   &   X           &   60 \\
    2013-05-14  &   56426   &   X           &   49 \\
    2013-05-17  &   56429   &   S,X         &   70,53 \\
    \tableline
    2013-05-23  &   56435   &   X           &   50 \\
    2013-05-30  &   56442   &   X           &   58 \\
    2013-06-21  &   56464   &   X           &   54 \\
    \tableline
    2013-07-14  &   56487   &   X           &   71 \\
    2013-07-15  &   56488   &   L,S         &   120,132 \\
    \textbf{2013-07-27}  &   \textbf{56500}   &   L,S,X       &   186,108,68 \\
    2013-07-28  &   56501   &   L,S*        &   133,117 \\
    2013-08-03  &   56507   &   L,S         &   112,75 \\
    \textbf{2013-08-12}  &   \textbf{56516}  &   L,S,X       &   120,60,56
    \enddata
    \tablecomments{The listed dates and MJDs for the epochs are representative of the majority of the epoch, not the start time; observations on the same day were taken in tandem.  The two boldfaced epochs are those for which we have simultaneous observations with \cxo.  The lower half (400~MHz) of the two S-band observations with an asterisk were corrupted and unusable.  The horizontal lines separate the epochs during which the three observed types of X-band profile are seen (see \S\ref{prof_var} and Figure~\ref{Xprofs}).}  
\end{deluxetable}

\subsection{X-ray}
\label{xray_obs}

\begin{figure}
\epsscale{1.0}
\plotone{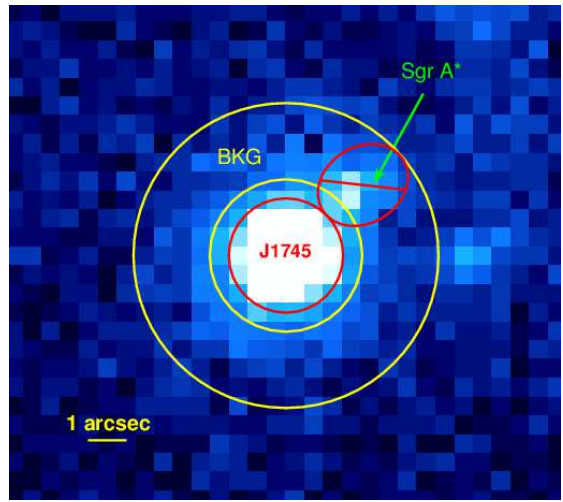}
\caption{\cxo\ field of \gcm\ for observation 15041.  1 ACIS pixel = $0.492''$.  The source counts were taken from the central-most encircled region (red circle).  Background counts were extracted from the annulus between the outer two (yellow) circles, excluding the area marked as ``Sgr\,A*''.  We account for pile-up as described in \S\ref{xray_obs}.}
\label{ds9}
\end{figure}

During two of our radio epochs, MJD~56500 and MJD~56516, we obtained simultaneous observations of \gcm\ with the \cxo\ X-ray Observatory (Obs. IDs 15041 \& 15042; PI: D. Haggard).  Table~\ref{xobs_table} contains details of the X-ray observations \citep[for further details see][]{Zelati15}.  The field of the first observation is shown in Figure~\ref{ds9}; the second observation was essentially the same.  In each observation, \gcm\ was positioned on the back-illuminated chip S3 of the Advanced CCD Imaging Spectrometer \citep[ACIS,][]{Garmire03} instrument.  The data were reprocessed with the \cxo\ Interactive Analysis of Observations software package \citep[\textsc{ciao}, version 4.6,][]{Fruscione06} and the calibration files in the \textsc{caldb} release 4.5.9.

In both observations, \gcm\ was bright enough to cause pile-up in the ACIS detector.  A ``pile-up map'' created with the \textsc{ciao} tool {\tt pileup\_map} confirmed that mild pile-up was present.  Exclusion of data near the center of the point-spread function (PSF) from the analysis would have resulted in the loss of too many photons (∼63\% of the source counts were in the two central pixels).  Moreover, the external part of the PSF contained a substantial number of counts from Sgr A*.  We thus decided to proceed as follows.

We extracted the source counts from a circular region centred on \gcm\ with a $1.5''$ radius (see Figure~\ref{ds9}); this region includes the piled-up events.  This area covers $\sim$85\% of the \cxo\ PSF (encircled energy fraction) at 4.5~keV.  A larger radius of 2--2.5$''$ would let in more counts from Sgr\,A* and would only marginally increase the encircled energy fraction.  Because of the complex environment, the background spectrum needed to be extracted close to the source.  We used a thin annulus (with radii of $2''$ and $4''$), excluding a bright area associated with Sgr\,A*.  The spectra, the ancillary response files and the spectral redistribution matrices were created using {\tt specextract}.  Following \citet{Rea13}, we adopt a pure blackbody for the spectral shape.  We corrected the spectra using the pile-up model by \citet{Davis01}, as implemented in the modeling and fitting package \textsc{sherpa} \citep{Freeman01}.  The pile-up fraction, estimated by fitting the {\tt jdpileup} model, is 3.7\% for the first observation, and 4.1\% for the second.  We did not attempt any correction of the light curves; the pile-up fraction is modest and, in general, pile-up affects spectra more than it does light curves and pulse profiles\footnote{This is true unless the pulse profiles are strongly dependent on energy, which is not the case for \gcm, though we refer the reader to \citet{Zelati15} for further details.}.  The spectral model fits were acceptable only when the pile-up model component was included.  A summary of the spectral fits is given in Table~\ref{specs}.

\setlength{\tabcolsep}{2pt}
\tabletypesize{\scriptsize}
\begin{deluxetable}{ccccc}
    \centering
    \tablewidth{0pt}
    \tablecaption{\label{xobs_table}Summary of Simultaneous \cxo\ Observations}
    \tablehead{\colhead{Obs. ID} & \colhead{Radio epoch} & \colhead{Exposure time} & \colhead{Net source}        & \colhead{RMS pulsed} \\
               \colhead{}        & \colhead{[MJD]}       & \colhead{[ks]}          & \colhead{counts [10$^{3}$]} & \colhead{fraction [\%]}}
    \startdata
    15041 & 56500 & 45.4 & 15.7 & $28.8\pm1.5$ \\
    15042 & 56516 & 45.7 & 14.4 & $28.9\pm1.8$
    \enddata
    \tablecomments{The 1$\sigma$ uncertainties for the RMS pulsed fractions were determined from Monte Carlo simulations (cf. \citet{Gotthelf99}).  By another measure, the pulsed fractions --- defined as the difference between the profile maximum and minimum divided by their sum --- are $\sim$48\%.  The folded profiles are shown in Figure~\ref{align}.}
\end{deluxetable}

\setlength{\tabcolsep}{5pt}
\tabletypesize{\scriptsize}
\begin{deluxetable*}{ccccccccc}
    \centering
    \tablewidth{0pt}
    \tablecaption{\label{specs}\cxo\ Spectral Results}
    \tablehead{\colhead{Obs.\,ID} & \colhead{$\mu\tablenotemark{a}$} & \colhead{$f\tablenotemark{a}$} & \colhead{\nh} & \colhead{$kT\tablenotemark{b}$} & \colhead{$R\tablenotemark{b}$} & \colhead{Observed flux$\tablenotemark{c}$} & \colhead{Luminosity$\tablenotemark{c}$} & \colhead{$\chi^2_{\textrm{red}}$ (dof)} \\
                                  & & & \colhead{[$10^{23}$ cm$^{-2}$]} & \colhead{[keV]} & \colhead{[km]} & \colhead{[$10^{-12}$ \flux]} & \colhead{[$10^{35}$ \lum]}}
    \startdata
    15041 & $0.50_{-0.05}^{+0.31}$ & 99.8\% & $1.26\pm0.03$ & $0.82\pm0.01$ & $2.34_{-0.17}^{+0.13}$ & $8.9\pm1.2$ & $3.2^{+0.4}_{-0.3}$ & 1.00 (288) \\
    15042 & $0.48_{-0.08}^{+0.25}$ & 97.1\% & $1.23\pm0.03$ & $0.83\pm0.01$ & $2.16_{-0.16}^{+0.13}$ & $8.1^{+1.4}_{-1.1}$ & $2.8\pm0.4$ & 1.00 (287)
    \enddata
    \tablenotetext{a}{Parameters of the {\tt jdpileup} \textsc{sherpa} pile-up model; $\mu$ is the grade-migration parameter and $f$ is the fraction of the PSF treated for pile-up, required to be in the range 85--100\%. For details, see \citet{Davis01} and ``The \cxo\ ABC Guide to Pileup''.}
    \tablenotetext{b}{The blackbody temperature and radius are calculated at infinity and assuming $D = 8.3$~kpc \citep{Genzel10}, which is assumed throughout this work.}
    \tablenotetext{c}{In the 0.3--8 keV energy range; for the luminosity we again assumed $D = 8.3$~kpc.}
    \tablecomments{The abundances used in the absorbed blackbody model are those of \citet{Anders89} and photoelectric absorption cross-sections are from \citet{Balucinska92}.  See \cite{Zelati15} for a complete treatment of these observations in the context of a long-term X-ray monitoring campaign.  Parameter uncertainties in the table are 1$\sigma$.} 
\end{deluxetable*}

\section{Results}
\label{results}

\subsection{Transient Events}
\label{trans}
\gcm\ is known to show narrow individual pulses \citep{Spitler14, Bower14, Lynch14}, similar to the radio magnetar J1622$-$4950 \citep{Levin12}.  We performed a cursory analysis of \gcm's individual pulses in our X-band data, seeking only to find anomalous burst-like events in the radio data that might be coincident or correlated with X-ray features or flares.  For this, we took two approaches.  In the first case, we folded the raw data into single-rotation integrations, approximately maintaining the original temporal resolution, averaging over frequency, and summing the polarizations.  These data were inspected visually.  In the second case, we analyzed the raw data with the \textsc{Presto}\footnote{\url{http://www.cv.nrao.edu/~sransom/presto/}} pulsar software package.  Here, we applied an RFI mask to the raw data with {\tt rfifind}.  We then made a dedispersed\footnote{We used the \citet{Eatough13b} DM of 1778~\dmunit, and compared the results to those from times-series dedispersed at 0~\dmunit\ and twice the nominal DM in order to discriminate between transient RFI and candidate pulses.}, frequency-averaged time-series with {\tt prepdata} for each X-band epoch, and searched for single pulses with the boxcar-convolution algorithm implemented in {\tt single\_pulse\_search.py}.  We repeated this process on the unmasked raw data.

Single pulses were detected; indeed, one to several pulses are visible by eye during almost every rotation.  However, we saw no anomalously large single pulses or other bursts.  The distributions of estimated single pulse energies all peak at $\lesssim$1 times the average profile energy and were inconsistent with power-law distributions.  The phases of the single pulse arrival times were consistent with occuring within the on-pulse window, and the distributions of the resolved single pulse widths peaked near 3-4 samples $\approx 2$~ms, in agreement with the X-band scattering timescales found by \citet{Bower14}.

Similarly, the (unfolded) X-ray light curves during the two simultaneous observations, binned from 0.5 to 5000~s, were featureless and constant.  The $\chi^2$ probability of constancy was high for both observations, regardless of the choice of binning ($>$30\% and frequently approaching 100\%).  Due to the uniformly poor quality of the X-band data as previously described, we refrain from further analysis or discussion of this aspect of \gcm\ and direct the reader to the observations of its X-band single pulses as observed with the Very Large Array and the GBT in \citet{Bower14} and \citet{Lynch14}, respectively.


\subsection{Profile Variability}
\label{prof_var}
Figure~\ref{Xprofs} shows examples of the three general types of observed X-band profiles, as well as corresponding examples of our baseline removal and on-pulse determination.  The transition between ``Type~1'', with a single main component having a trailing-side shoulder and a more quickly rising leading edge, and ``Type~2'', with the main component having a leading-side shoulder and a nub feature on the trailing side, happens more than three weeks after the X-ray burst on MJD 56407 and more than two weeks before the burst on MJD 56450.  The ``Type~1'' shape was seen as early as a week after the discovery \citep{Eatough13a} and published in \citet{Eatough13b}.  Similarly, the transition between ``Type~2'' and ``Type~3'', which has a larger two-peaked component, happened more than two weeks after the burst on MJD 56450 and more than three weeks before the burst on MJD 56509.  For these reasons, we do not associate the profile types (which are most likely not absolutely discretized) with the observed X-ray bursts.  Within a single observation, the profile shape did not change between 5 min subintegrations.

\citet{Lynch14} also documented the time-variability of \gcm's X-band profile as seen with the GBT.  As their first observation is coincident with our last observation, they have also seen the ``Type~3'' shape, which persists and evolves during most of what they have labeled a ``stable~state''.  This ``stable~state'' is characterized by relatively smooth profile transitions, a gradual flux evolution, and a phase-connected timing solution --- all in contrast to what they call an ``erratic~state'', which is onset sometime after MJD~56682.  Later in their observations, during epochs with MJDs~56794 and 56865 (both in the ``erratic-state''), they see a profile resembling what we have labeled ``Type~1''.  We note that we did not witness any of the very sporadic profile variability seen in \citet{Lynch14} associated with the ``erratic~state'' (e.g., the drastic profile changes seen in their last two observations, separated by only eleven days), but rather we observed each of these three types only for a single interval of time.

\subsection{Timing}

\label{timing}

Between having bursts, glitches, unstable profiles, and timing noise, magnetars are notoriously some of the hardest pulsars to time (cf. the original radio magnetar J1809$-$1943 \citep{Camilo06}, or see a recent review of magnetars in \citet{Mereghetti13}).  As is evident from the X-ray and radio timing in \citet{Zelati15}, \citet{Lynch14}, and \citet{Kaspi14}, obtaining a single phase-connected timing solution for \gcm\ is difficult, due to a significant level of timing noise.  Here, we measure an overall average spin-down for the purpose of summing the data in each epoch.

Pulse times-of-arrival (TOAs) were measured by cross-correlating the time- and frequency-averaged data profiles with smoothed, ``noise-free'' template profiles using standard \textsc{PSRCHIVE} routines.  The templates are generated by arbitrarily aligning and averaging all of the data for which the template is used.  Single templates were used for the L- and S-band data, but three separate templates were used for X-band, depending on the profile observed, as discussed in \S\ref{prof_var}.  Arbitrary phase offsets were fit between TOAs measured from all of the different templates as part of the timing models.  These phase offsets serve to align the template profiles, but do so indiscriminately with respect to pulse broadening from interstellar scattering; this has the effect of biasing DM estimates if one tries also to measure the dispersive delay between TOAs of different frequencies.  See \S\ref{wideband_dms} for our DM measurements based on wideband modeling of the L- and S-band data.

\begin{figure}
\epsscale{1.0}
\plotone{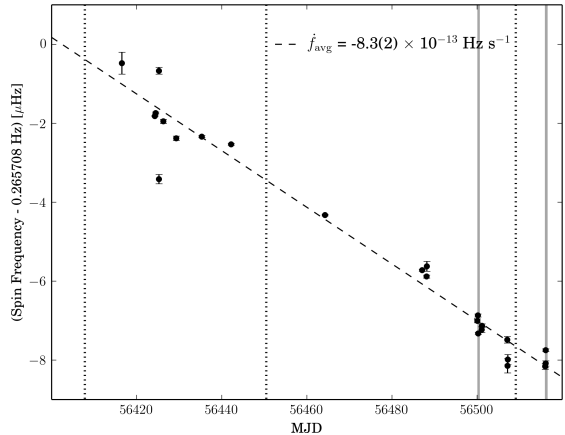}
\caption{Average spin evolution of \gcm.  The three vertical dotted lines correspond to the three X-ray bursts detected by \swift/BAT.  The two vertical grey bars cover our \cxo\ observations.  Measurements from the two early S-band observations are not included, nor from the X-band epoch on MJD 56425, as they were very significant outliers.  The quoted uncertainty does not include the residual scatter.}
\label{spin_evol}
\end{figure}

Figure~\ref{spin_evol} shows the measured values of the spin frequency $f$ as a function of time.  The average measured spin-down of $\dot{f}_{avg} = -8.3(2) \times 10^{-13}$~Hz~s$^{-1}$ was sufficient to average the data in each epoch with negligible smearing for the flux measurements (\S\ref{flux}), and is a reasonable approximation for the overall trend in the spin evolution\footnote{Quantities in parentheses represent the 1$\sigma$ uncertainty on the last digit in the respective measurement throughout the paper.}.  This average value also lies between the two $\dot{f}$ values presented in Table~2 of \citet{Kaspi14} for the same range of dates.

Although we are not interested in a full timing solution for these data in this work, we found corroborative results when following the suggestion in \citet{Kaspi14} that there is an abrupt change in $\dot{f}$ around the time of the \swift/BAT-observed X-ray burst on MJD 56450.  Namely, while a single, predictive timing solution was not found, our pre- and post-burst TOAs are described by two simple phase-coherent solutions with parameters $\dot{f}_{\textrm{pre}} = -5.005(1) \times 10^{-13}$~Hz~s$^{-1}$, $\dot{f}_{\textrm{post}} = -9.4799(5) \times 10^{-13}$~Hz~s$^{-1}$, and $\ddot{f}_{\textrm{post}} = -2.696(6) \times 10^{-20}$~Hz~s$^{-2}$.  These values are in good agreement with those in \citet{Zelati15}, \citet{Kaspi14}, and \citet{Rea13}, although we were not sensitive to $\ddot{f}_{\textrm{pre}}$.   We could only obtain a single phase-connected timing solution for all of the TOAs by using five spin frequency derivatives, which is not a predictive ephemeris.

\subsubsection{Profile Alignment}
\label{alignment}

\begin{figure*}
\epsscale{1.0}
\plotone{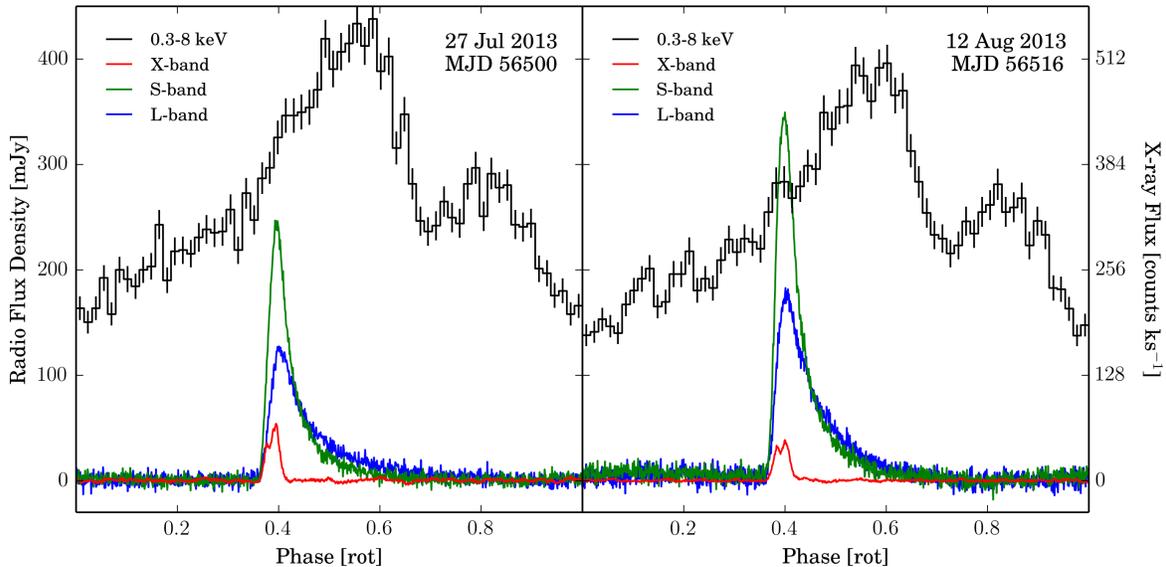}
\caption{Absolute phase alignment of \gcm's radio and X-ray profiles determined separately on two days.  Note that the brightest radio profile is seen in S-band (see Figure~\ref{spectrum_56516}).  The profiles have 1024 and 64~phase~bins, respectively, and are shown as they would be observed at the Solar System barycenter for phase-zero MJDs 56499.98000761 and 56515.96999979, referenced to infinite frequency.  The assumed dispersion measure is 1778~\dmunit.  During two later \xmm\ observations (presented in \citet{Zelati15}), there is a peculiar, narrow feature seen in the otherwise broad X-ray profile near the phase of radio emission as shown here (see text).}
\label{align}
\end{figure*}

In Figure~\ref{align}, we present the absolute alignment between the \cxo\ 0.3--8~keV X-ray profiles and the GBT radio profiles in L-, S-, and X-bands.  We determined an independent ephemeris for each of the two epochs from the radio data by fitting TOAs from each day for only the spin frequency, fixing the spin-down parameter at the average value reported above.  These TOAs were measured from the frequency-averaged data with 5 min subintegration resolution.  The phase-zero time was referenced to the arrival of infinite-frequency radiation at the Solar System barycenter, which assumes a constant dispersion measure of 1778~\dmunit\ between the two observations.  The X-ray photon arrival times were barycentered also using the sky position given in \S\ref{radio_obs} and the JPL Planetary Ephemeris \textsc{de-405}.  These events were folded into pulse profiles with 64~phase bins using the corresponding epoch-specific ephemeris by the {\tt prepfold} program of \textsc{Presto}.  The alignment based on folding using a single ephemeris for both epochs --- either the post-burst or the multiple frequency derivative ephemeris --- yielded indistinguishable results.  This is reasonable, since the RMS timing residual from either of those ephemerides is on the level of individual bins.  On the other hand, it may be surprising that there seemed to be no interruption in the ``post-burst'' ephemeris; the third detected X-ray burst occurred at the midpoint between the two simultaneous radio/X-ray epochs.

We modeled each of the two X-ray profiles with four Gaussian components to measure the relative offsets with respect to the radio profiles.  The offsets and their uncertainties were determined from Monte Carlo trials, where ``offset'' here refers to the phase that maximizes a cross-correlation such as the one prescribed in \citet{Taylor92}.  There was a small offset between the X-ray models, $\lesssim$0.02~rot.  A difference in DM would shift the relative phase between the X-ray profile and the S-band profile (our fiducial profile) only by $\sim$3$\times 10^{-4}$~rot per unit DM~[\dmunit].  Even for the DM difference of $\sim$17~\dmunit\ measured between these epochs (see \S\ref{wideband_dms} and Figure~\ref{LS_stuff}), the phase difference is $\sim$0.005~rot.  The remaining offset can be explained by a combination of the variability of the X-ray profile and timing noise, with the former being dominant.  After removing this difference, the offsets with respect to the radio profiles do not change between the two days within the variance of the measurements.  The phase offset relative to the S-band profile is approximately 0.15(1)~rot.   The radio magnetars J1809$-$1943 and J1550$-$5418 both also show rough alignment of pulsed radio emission with their X-ray profiles \citep{Camilo07a,Halpern08}, whereas no pulsed X-ray emission has been detected from J1622$-$4950 \citep{Anderson12}.

The two double-peaked X-ray profiles appear essentially featureless.  The RMS pulsed fractions are given in Table~\ref{xobs_table}.  There are not sufficient data to decompose the profiles into energy bands to look for meaningful spectral dependencies, although we wish to point out a possible transient feature that appears in the \xmm\ data recently published by \citet{Zelati15}.  In the energy-dependent \xmm\ profiles of Figure~4 from \citet{Zelati15}, there is a conspicuous narrow feature on the leading edge of the double-humped X-ray profile that is close to the phase of radio emission (within $\sim$0.05~rot).  It appears most prominently around phase $0.55$ in the 0.3--3.5~keV profile of the third \xmm\ observation (with Obs. ID 0724210501).  It is also seen in two of the other three energy-dependent profiles (except for the highest energy 6.5--10.0~keV profile), contributing to the integrated flux in the energy-averaged profile.  A similar feature is seen at the same phase in the first \xmm\ observation (with Obs. ID 0724210201) to a lesser extent.  According to their table, these observations were separated by 23 days, with the first occurring 19 days after the \cxo\ observations presented here (which are also included in \citet{Zelati15}).  The three \cxo\ observations and the one \xmm\ observation taken during these 23 days show no obvious feature, despite covering the same range of energies, although \cxo\ recorded only between 10 and 50\% of the counts as did by \xmm.  Therefore, without additional observations, it remains only a peculiarity.

\subsection{Radio Flux Density}
\label{flux}

From the radio data, we made measurements of \gcm's flux density as a function of time and frequency.  We measured the mean flux densities in 50~MHz wide channels and used a weighted average of these measurements to obtain representative flux densities for each band, per epoch.

For all of the L- and S-band profiles, we defined ``on-pulse'' regions as follows.  A model pulse profile for each frequency was determined from the wideband modeling described in \S\ref{wideband}.  We then found the smallest range of pulse phases that contained 99\% of the integrated flux density of the model profile.  Examples of the on-pulse windows for the scattered L- and S-band profiles can be seen in Figure~\ref{LSprofs}.  The mean flux density was calculated by averaging the observed flux density in the window and scaling it by the duty cycle.  The uncertainties were estimated by measuring the mean noise level in the last quarter of each profile's power spectrum\footnote{This is a robust method to estimate the off-pulse variance, assuming the profile is resolved \citep[e.g., see][]{Demorest07}.}.  We accounted for systematics in the residual profile by adding the scaled, residual mean flux density to the uncertainty in quadrature.  These corrections were small, as the reduced $\chi^2$ values of the residuals were usually $<$1.5 and always $<$2.0.

The measurement of the X-band flux densities was complicated by the dynamic baseline variations mentioned in~\S\ref{radio_obs}, as well as the intrinsic variability of the profile shape.  We used polynomial functions to remove the baseline variations on a profile-to-profile basis (e.g., see Figure~\ref{Xprofs}).  For these profiles, we first centered each profile to be near phase 0.5 to avoid edge-effects of the polynomial fit from affecting the on-pulse region.  A high degree polynomial function was fit to the baseline of each profile, where in the first iteration an on-pulse window with a duty cycle of 6\% was blanked out from the fit to avoid initially over-estimating the noise\footnote{None of the profiles had a smaller duty cycle than 6\% and a polynomial of degree 15 was used; this was the smallest degree polynomial that reasonably and automatically removed systematic baseline trends from all of the profiles without having to also vary the degree of the polynomial on a profile-to-profile basis.}.  The level of the residual off-pulse noise was calculated, and then the on-pulse window was widened until the flux density at the edges of the on-pulse region dropped below the noise level.  The baseline polynomial was then refit to the original profile, but with the new on-pulse window blanked out.  The mean flux density and its uncertainty were calculated in these baseline-removed, on-pulse windows as described for the lower frequency data above, but a systematic error was added in quadrature to the uncertainty that represented the mean flux density across the on-pulse phase window removed by the polynomial fit.  This tested method gives dependable, conservatively estimated X-band flux densities.

\subsubsection{Flux Evolution}
\label{fluxevol}

The radio flux evolution of \gcm\ is shown in Figure~\ref{flux_evol}.  The mean X-band flux density increases rapidly in the first half of our observations, increasing by at least a factor of $\sim$6 over fifty days, and then tapers off at the 1~mJy level.  The earliest reported measurement of \gcm's X-band flux density was $\sim$0.2 mJy, taken with the Effelsberg 100-m Radio Telescope, consistent with our GBT measurement two days later \citep{Eatough13a}.  Our data show a similar increase in the low frequency flux densities.  The S-band flux increases by about an order of magnitude over ninety days, and in our last five observations covering about thirty days, the average L- and S-band fluxes increase by a factor of two.  Given the measured scattering timescales for \gcm\ (see \S\ref{wideband_stuff}) and the recently measured proper motion of the pulsar, the timescale for refractive scintillation to be important is much larger than the span of our observations (see \citet{Bower15} for further discussion).

Having picked up where we left off, \citet{Lynch14} increased the cadence of GBT X-band observations after MJD~56516 and found a similar, slow increase of the flux, up to $\sim$3~mJy, over the next 170~days.  As already mentioned, after this ``stable~state'' of slow, steady flux increase, the authors found that \gcm\ entered an ``erratic~state'', characterized in part by a larger and highly variable X-band flux, similar to what was seen in two other radio magnetars \citep{Camilo07a,Levin12}.  Superimposed on top of this radio flux evolution is a relatively slow decay of the X-ray flux, compared to other magnetars \citep{Rea13,Kaspi14,Lynch14,Zelati15}.  Between our two simultaneous GBT/\cxo\ observations separated by $\sim$15 days, the radio flux increased by $\sim$60\% while the X-ray flux decreased by $\sim$10\%.  This trend (seen here and in \citet{Lynch14}) is opposite to those of the other radio magnetars, which show decreasing radio and X-ray flux with time over the course of an outburst \cite{Rea12a}.

\subsubsection{Radio Spectral Index}
\label{spec_index}

\begin{figure*}
\epsscale{0.8}
\plotone{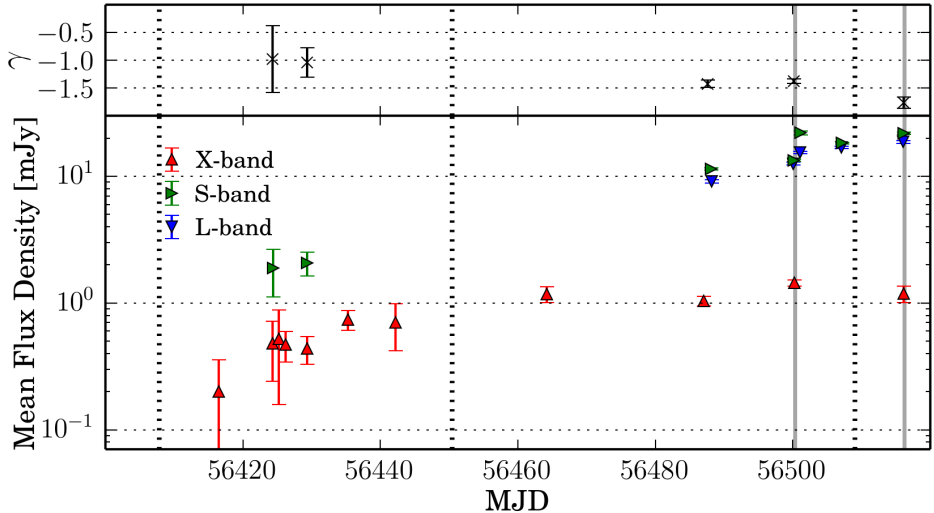}
\caption{The early radio flux (bottom panel) and spectral (top panel) evolution of \gcm\ over 100 days from the observations in Table~\ref{obs_table}.  The vertical demarcations are the same as in Figure~\ref{spin_evol}.  \citet{Lynch14} find a continuation of the slow, steady increase in X-band flux for another six months, which is followed by what they call an ``erratic~state''.  The apparent excess average S-band flux density during MJD~56501 is explained by the fact that the lower half of the band was corrupted (see Table~\ref{obs_table}), and the pulsar's flux density apparently increases with frequency in this range (see Figure~\ref{spectrum_56516}).  The average value of the spectral index $\gamma$ is about $-1.4$; see text for details.}
\label{flux_evol}
\end{figure*}

Because we have essentially simultaneous observations\footnote{In one case, the X-band observation was taken a day earlier; see Table~\ref{obs_table}.} of \gcm\ in frequency bands spaced by two octaves, we can measure the spectral index $\gamma$, where $S_{\nu} \propto \nu^\gamma$ for flux density $S_\nu$ at frequency $\nu$.  The upper panel of Figure~\ref{flux_evol} shows $\gamma$ as measured between the average X-band flux density and the combined average flux densities of the lower frequency band(s).  The error bars were approximated by varying the average fluxes within their measurement uncertainties.  The decorrelation bandwidth for diffractive scintillation is much smaller than even our native frequency resolution and will not be a source of variability here.

There is no large, obvious stochasticity, as opposed to, for example, J1809$-$1943 \citep{Lazaridis08}, but there may be a trend.  \citet{Shannon13a} report two early measurements of $\gamma$ across the bands spanning 4.5--8.5~GHz and 16--20~GHz.  The first measurement on MJD~56413 is close to $-1.0$ in the high frequency band, though it is closer to 0.0 in the lower frequencies, and the second on MJD~56443 is $\sim -1.0$ across both bands, consistent with our measurements more than two weeks prior.  Our three later measurements indicate a significantly steeper spectrum.  The average value for $\gamma$ of $-1.4$ is tantamount to the average spectral index for normal pulsars across gigahertz frequencies as reported in \citet{Bates13}.  \citet{Camilo07b} and \citet{Anderson12} both make mention of a general steepening of the spectral indices of J1809$-$1943 and J1622$-$4950, respectively, despite remaining much flatter than what is seen in \gcm.  However, \citep{Lazaridis08} finds the opposite for J1809$-$1943 in later observations.

This finding apparently breaks the mold set by the other three radio magnetars, which have essentially flat (or inverted) spectra \citep{Camilo06,Camilo08,Levin10,Keith11}.  However, no firm conclusions can be drawn from this handful of measurements from early times in \gcm's outburst, especially knowing that the other radio magnetars also show a variable radio spectrum \citep{Camilo07b,Lazaridis08,Anderson12}.  In fact, at the time of writing, the findings of \citet{Torne15} suggest that at much later times (a year after the present observations), the radio spectrum of \gcm\ between 2 and 200~GHz was much flatter, with $\gamma = -0.4(1)$.

\subsubsection{Spectral Shape}
\label{spectrum}

\begin{figure}
\epsscale{1.0}
\plotone{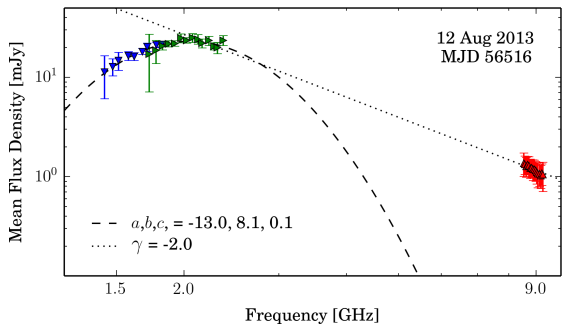}
\caption{An example of \gcm's radio spectrum from the brightest observed epoch, MJD 56516.  The markers are as in Figure~\ref{flux_evol}; note that the flux densities agree in the $\sim$100~MHz overlap between L- and S- band.  A similar inverted parabolic shape over log-frequency is seen during the other sets of (nearly) simultaneous observations, which is reminiscent of the so-called GPS pulsars.  The coefficients $a$, $b$, and $c$ of the fitted dashed parabola (log$_{10}$(S$_{\nu}$) = $ax^{2}+bx+c$, for x = log$_{10}$($\nu$)) are given in the plot, along with the spectral index $\gamma$, which for this plot was fitted between the \textit{peak} of the parabola and the X-band data.}
\label{spectrum_56516}
\end{figure}

One example of \gcm's radio spectrum is shown in Figure~\ref{spectrum_56516}; the spectra from the other days are qualitatively similar.  The spectrum shows a non-power-law increase in flux between 1.4 and 2.4~GHz, with a possible peak near 2~GHz.  The inverted log-parabolic shape is reminiscent of what have been called ``gigahertz-peaked spectra'' (GPS) pulsars \citep{Kijak11,Kijak13,Dembska14,Dembska15}, although the GPS pulsars supposedly have a much broader spectral shape, over a dex in frequency.  For reference, we fit a log-parabola to the low frequency points, the parameters of which are given in the figure.

It is difficult to explain the spectral shape we see in the lower frequencies.  It is conceivable that the dense, unique environment near \gcm\ in the Galactic Center significantly alters the spectral shape of radio emission between 1 and 10~GHz (e.g., via free-free absorption, although the detection of Sgr~A* at 330~MHz implies a low free-free optical depth of $\lesssim$1 \citep{Nord04}), but it is difficult to draw any conclusions without a dedicated set of observations.

Another possibility is that we have systematically under-estimated the flux: one well known source of bias comes from under-estimating the flux at low frequencies due to significant area in the scattering tails being lost in the calculation of the baseline flux.  However, even at 1.4 GHz the scattering timescale is $\sim$500~ms $\approx$ 0.13~rot (see \S\ref{wideband_stuff}).  In the worst case of a Kolmogorov scattering index ($-4.4$), the scattering timescale at our lowest frequency is no more than $\sim$20\% of a rotation.  As mentioned in \citet{Kijak11} and treated graphically in \citet{Macquart10}, the pulsed fraction drops by only $\sim$10\% when the scattering timescale is \textit{half} the pulse period.  Therefore, we can suggest that at worst we are underestimating the L-band flux densities at the $\sim$10\% level, but this still would imply a positive or approximately flat spectral index between 1.4 and 2.4~GHz; the observed flux density drops precipitously somewhere thereafter.

A more promising, albeit provisional possibility has been offered up by recent modeling of the \citet{Shannon13a} observations. \citet{Lewandowski15b} make a case study of \gcm\ to demonstrate the possibility of thermal free-free absorption as the explanation for the GPS.  For \gcm, the authors suggest a combination of an expanding ejecta and/or an external absorber to explain the changing spectrum seen early after the initial outburst in \citet{Shannon13a}.  The free-free absorbed model spectra offer a reasonable explanation for the lack of low-frequency detections of \gcm\ immediately after the initial outburst and detections above 4~GHz; our two early S-band observations may support this idea.  Our spectra from three months later may also inform the story of an evolving or endemic free-free absorbing medium in the environment of \gcm.

\subsection{Wideband Portrait Model}
\label{wideband}

As is evident from Figure~\ref{LSprofs}, \gcm\ has a highly scattered, simple profile across a gigahertz bandwidth, from 1.4 to 2.4 GHz.  For a nominal DM value of 1778~\dmunit, there is a delay of $\sim$0.66 rotations across this band, which is easily measurable.  All of the average L- and S-band profiles showed prominent scattering tails from multipath propagation through the interstellar medium (ISM).  The quality of the data permitted us to make ``wideband'' measurements of both the DM and the scattering timescale~$\tau$, as well as its power-law index $\alpha$, on an epoch-to-epoch basis\footnote{The two earliest S-band observations were exceptions; corrupted data, low signal-to-noise ratios, and the lack of L-band data resulted in uninformative measurements of the DM, $\tau$, and $\alpha$.  This is also the reason these observations were excluded from the average $\dot{f}$ measurement in \S\ref{timing}.}.

For this, we used the methods and augmented software described in \citet{PDR14} to make a wideband ``portrait''\footnote{We use the word ``portrait'' to mean the total intensity profile as a function of frequency.} model for each of the five epochs where we have both L- and S-band observations.  For each of these epochs we combined the data from the two low frequency bands in a fit for a global portrait model that included a single scattered Gaussian component with profile evolution parameters, a constant baseline term, a phase offset between the bands, DMs for each band, and the scattering index $\alpha$.  The scattering timescale is defined in the usual way by assuming a one-sided exponential pulse broadening function for the ISM, so that an observed profile $p(\varphi)$ is the convolution given by
\begin{equation}
\label{Pscat}
p(\varphi) = g(\varphi) \ast e^{-\frac{\varphi P_s}{\tau}}\textrm{H}(\varphi),
\end{equation}
where $\varphi$ is the rotational phase, $P_s$ is the spin period, H is the Heaviside step function and $g(\varphi)$ is the intrinsic total intensity profile shape.  For a power-law spectrum of density inhomogeneities in the ionized ISM $\tau$ is expected to have a power-law dependence on frequency $\nu$ as
\begin{equation}
\label{tau_eq}
\tau(\nu) = \tau_{\nu_\circ}\big(\frac{\nu}{\nu_\circ}\big)^\alpha,
\end{equation}
with reference frequency $\nu_\circ$.  In all cases, the scattering timescale (133~ms at 2~GHz; see below) dominates the smearing from the process of incoherent dedispersion ($\sim$0.7~ms ($\nu$/2~GHz)$^{-3}$), the smearing from an incorrect DM when averaging channels ($\sim$25~$\mu$s ($\delta$DM/\dmunit) ($\nu$/2~GHz)$^{-3}$),  and the temporal resolution (1.8~ms for 2048 profile bins), so we have not included those modifications of the pulse profile shape in the model.  However, deviation from the simple timing models discussed in \S\ref{timing} (e.g., see Figure~\ref{spin_evol}) during any of these epochs could add profile smearing in the integrated profiles (at the level of $\sim$tens of ms --- a significant fraction of the scattering timescale).  We avoided this source of bias by iterating over the timing model to remove the timing residual on a per-epoch basis.

We model $g$ with a frequency-dependent Gaussian function,
\begin{equation}
\label{gauss}
g(\nu,\varphi) = A(\nu)~\textrm{exp}\big(-4\textrm{ln}(2)\frac{(\varphi-\varphi_g(\nu))^2}{\sigma(\nu)^2}\big),
\end{equation}
which is parameterized by its location $\varphi_g(\nu)$, full-width-at-half-maximum (FWHM) $\sigma(\nu)$, and amplitude $A(\nu)$.

As described in \citet{PDR14}, each of these parameters nominally has an additional parameter describing its frequency dependence.  However, because this combined band has a fractional bandwidth of ``only'' $\sim$0.5, we assume $\varphi_g(\nu) = \varphi_\circ$ is a frequency-independent value.  That is, we assume there is no drift intrinsic to the one component across the band.

Furthermore, when allowing for a frequency-dependent $\sigma$, we found no significant evolution, and so we chose also to fix the evolution $\sigma(\nu) = \sigma_\circ$ to be a frequency-independent fit parameter in our final portrait models.  This choice was further justified by performing independent per-channel profile fits of a single, scattered Gaussian component and examining the frequency evolution of $\sigma$.  Also, there are X-band observations for three of these epochs, and in these cases the FWHM of the X-band profiles (all of ``Type 3''), when fitted with a single, unscattered Gaussian component, was always within the scatter of those measured from the lower frequency observations.  These results are consistent with the weak (or lack of) frequency dependence of $\sigma$ found in \citet{Spitler14}.

We normalized the intensities of the data to be fit by the maximum profile value in each frequency channel to remove the unusual spectral shape (see \S\ref{spectrum}).  This allowed the Gaussian amplitude to be easily modeled by a power-law function for $A(\nu)$.  In all cases, the reduced $\chi^2$ of the fit was $<$1.1, and a second Gaussian component was never justified by the residuals.

The combination of the quality of the X-band data, the variability of the profile, and the expected value of $\tau$ at 8.9~GHz ($\lesssim$1 ms, comparable with our native time resolution) was such that we did not attempt to incorporate this high frequency data into our wideband profile model, nor did we measure the scattering timescale in either the average profile or the single pulses.  We refer the reader to \citet{Bower14} and \citet{Spitler14} for high frequency scattering measurements of \gcm.

\subsubsection{Pulse Width \& Scattering Parameters}
\label{wideband_stuff}

\begin{figure}
\epsscale{1.0}
\plotone{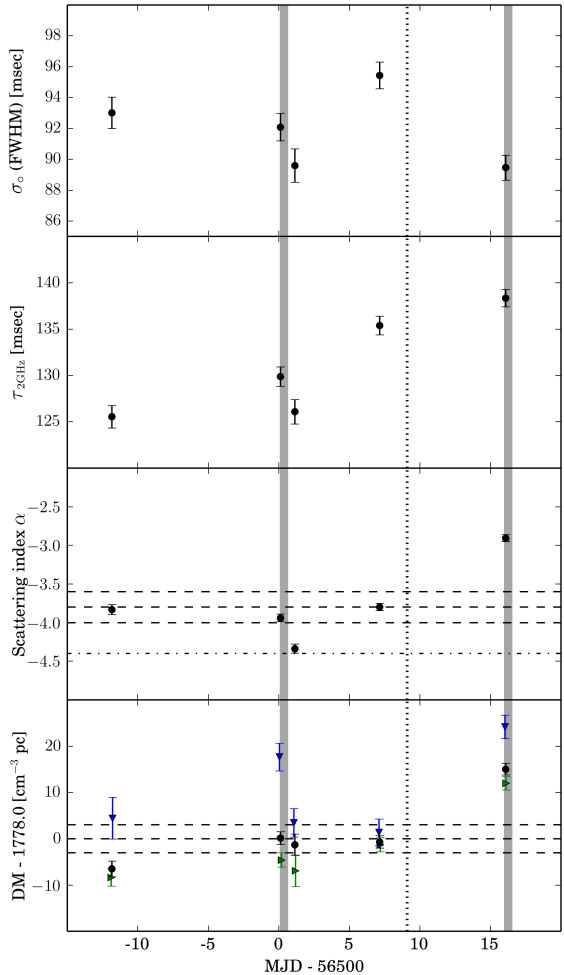}
\caption{Our wideband measurements from five epochs.  The vertical demarcations are the same as in Figure~\ref{spin_evol}.  The FWHM showed neither frequency nor temporal dependence.  The trend in the scattering timescale $\tau$ is less scattered and more precise than the measurements presented in \citet{Spitler14}.  The dash-dotted line in the panel for the scattering index $\alpha$ marks the fiducial Kolmogorov value of $-4.4$, and the dashed lines mark the \citet{Spitler14} measurement of $-3.8$(2).  The additional markers in the bottom panel are the same as in Figure~\ref{flux_evol}: the blue/down-pointing triangles are L-band measurements, and the green/right-pointing triangles are S-band measurements --- the dots are their weighted average.  The dashed lines here are the \citet{Eatough13b} DM of 1778(3)~\dmunit.  See the text for a discussion of the DM measurements.}
\label{LS_stuff}
\end{figure}

The results from our wideband models are shown in Figure~\ref{LS_stuff}.  There was no significant change in the measured FWHM of the unscattered profile, and our average (frequency-independent) value of 91.9(4)~ms~=~0.0244(1)~rot is also consistent with what is reported in \citet{Spitler14}.  The scattering timescale at 2~GHz, $\tau_{2\textrm{GHz}}$, appears to increase by $\sim$10\% over the four weeks, and the scattering index $\alpha$ deviates from its average value, first to a Kolmogorov value near $-4.4$ (the dash-dotted line), and then to a much shallower value near $-3.0$.  Both of these results are somewhat peculiar, but similar variations are also reported in \citet{Spitler14}, though they do not discuss the temporal evolution of either quantity.  That is, their published values of $\tau$ from a variety of epochs and frequencies cannot be unified by a single scattering timescale and index.  In fact, their measurements of $\tau$ show more scatter over the course of their observations than those presented here, which have some overlap.  When the authors combine all of their measurements, they find an average value for $\alpha$ of $-3.8$(2).

\begin{figure}
\epsscale{1.0}
\plotone{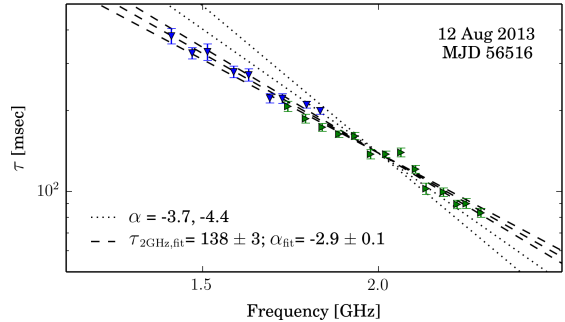}
\caption{Independent per-channel measurements of the scattering timescale $\tau$ and the fitted scattering index $\alpha$ for the brightest set of L- and S-band observations, on MJD 56516; similar plots from the other four days have a significantly more negative slope.  The dashed lines represent our measurement, whereas the dotted lines show our average value of $\alpha$ from wideband modeling and the fiducial Kolmogorov value of $-4.4$, all with the same value of $\tau_{\textrm{2GHz}}$.  Here, the measurement uncertainties have been inflated by the reduced $\chi^2 \sim$ 2.}
\label{scat_56516}
\end{figure}

We checked our measurements in two ways.  First, we performed conventional profile fits of a single, scattered Gaussian component to each individual frequency channel, independent of any evolutionary constraint.  The values of $\sigma$, $\tau$, and $\alpha$ for each epoch were consistent with what we found by applying the wideband modeling method.  In Figure~\ref{scat_56516}, we show the measurements of $\tau$ measured in this way for the brightest observed epoch (MJD~56516) and over-plot the fitted power-law, which has the most extreme $\alpha$ value of the five epochs.  There was nothing unusual about the data from this epoch in terms of RFI, data removal, calibration, or baseline variations.  Second, as a check for our average values, we summed all of the data portraits together by coherently stacking the observations (having fit for a phase and DM in each epoch), and fit a single wideband model to the averaged data (with the same constraints as earlier).  Using this method, we obtained similar average values: $\sigma = 0.0246(1)$~rot, $\tau_{\textrm{2GHz}} = 133.0(5)$~ms, and $\alpha = -3.71(2)$, the latter of which is in concert with the average $\alpha$ value from \citet{Spitler14}.  Our extrapolated value of $\tau_{\textrm{1GHz}} = 1.74(3)$~s is only slightly at odds with their average value of  $\tau_{\textrm{1GHz}} = 1.3(2)$~s, which is probably due to the temporal variability of $\tau$.  As others have noted (e.g., \citet{Bower14}), the anticipated value for $\tau_{\textrm{1GHz}}$ along this line of sight based on empirical relationships, for a DM of 1778~\dmunit, is about 600$\times$ larger than what is observed \citep{Krishnakumar15,Lewandowski15a}.

\citet{Bower14} determined $\tau_{\textrm{8.7GHz}} \lesssim$ 2~ms from interferometric measurements of \gcm's single pulses, implying that a value for $\alpha$ as shallow as $-3$ is not unbelievable.  Furthermore, scattering measurements from two high-DM pulsars discovered near the Galactic Center (both within 0.3\degr\ and having DMs 1100--1200~\dmunit) implied $\alpha = -3.0(3)$ \citep{Johnston06}.  It is not uncommon for pulsars to have $\alpha > -4$, particularly along special lines-of-sight, and it is empirically suggested that the highest DM pulsars may have an average scattering index significantly shallower than $-4$ \citep{Lohmer01,Lohmer04,Lewandowski15a}.  Note that observing $\alpha \ne -4.4$ does not necessarily imply a non-Kolmogorov spectrum of density inhomogeneities; rather, it could be that a non-thin-screen geometry may be responsible \citep{Cordes01,Lewandowski13}.

\subsubsection{Dispersion Measures}
\label{wideband_dms}

The bottom panel in Figure~\ref{LS_stuff} shows the best-fit DMs as determined by the wideband models in the essentially simultaneous L-band and S-band observations (blue/down-pointing and green/right-pointing triangles, respectively).  The black points are the weighted average of the two measurements; there is obviously some variance about the nominal value of 1778(3)~\dmunit, and our overall average value is $\sim$1781(1)~\dmunit.  Without exception, the measured L-band DMs are greater than those measured in S-band.  There is also only one epoch where the 1$\sigma$ uncertainties have any overlap; the RMS variance of the differences is $\sim$6~\dmunit.  The absolute DM differences cause residual dispersion on the order of $\lesssim$20~ms $\sim$ 10~bins (for 2048-bin profiles) across the corresponding band, and so they present significant profile deviations.  To make sense of the discrepant DMs between the two frequency bands, we consider a number of possibilities.

First, the time-averaged data for each epoch showed few systematics with negligible baseline variations, so we do not believe that data quality was an issue here.

Next, as is well known, the measured absolute DM will be affected by the choice of profile alignment\footnote{For example, DMs are significantly biased when either assuming a constant profile shape in the presence of scattering, or aligning scattered profiles by conventional methods because the convolution of the ISM pulse broadening function with a profile of finite width introduces a delay that is a function of the scattering timescale (i.e. frequency).  This is partly why proper wideband modeling is necessary.}.  We can rule out any simple, constant profile evolution as the source of the differing DMs because such a modification introduces a \textit{constant} difference in the DMs; the changes in the measured DM should be the same independent of the choice of alignment.  Even if our assumption that there is no intrinsic drift in the location of the (unscattered) profile component across the band is wrong, allowing for a drifting component will still reproduce discrepant DMs; we have verified this by allowing for frequency evolution in the location parameter of the Gaussian component.

A second confounding element from our modeling could be the use of different models for each epoch; if they are all systematically wrong in their alignments or representation, they could be wrong differently.  One way to check this is to simply use one fixed model to remake the DM measurements.  We used the average portrait model discussed earlier and confirmed that the DMs remain similarly extreme, within $\sim$2~\dmunit, comparable with the measurement uncertainties.  In fact, we tried a large number of fixed and variable portrait models, but never obtained either consistent DMs or DMs with a near constant offset.  So, to the extent that $\tau$ and/or $\alpha$ are measurably changing, we are justified in keeping them as free parameters for each epoch's model.

Similarly, the known profile variability that is seen in all of the radio magnetars could also play a role when using either a fixed or variable portrait model.  However, besides the flux density, any underlying profile shape changes either with time or frequency are masked by the large level of scattering.  As mentioned, the FWHM does not seem to change significantly in either time or frequency.  Furthermore, the three X-band observations taken during these epochs show no large profile changes, and are all of the ``Type~3'' shape.

Next we can ask whether or not the slight asynchronicity could have any effect; that is, could the DM change so significantly on $\sim$hour timescales?  We will return to this question below, but it is not an uncommon \textit{a priori} assumption to expect that the observed DM does not change between observations separated by $\lesssim4$~hr.

One could also ask if the method by which we measure the DMs introduces a systematic error, where the error may depend on the exact values of $\tau$ and $\alpha$, or even the spectral shape.  To answer this, we performed a number of Monte Carlo simulations.  In the simulations, we used the models from MJDs 56500 and 56516, which have the most extreme values for the difference between the DMs, and the most extreme values for $\tau$ and $\alpha$, respectively.  For each trial, we made fake L- and S-band observations by appropriately constructing the model for that band and scaling each frequency channel's amplitude to match the spectral shape.  We then added random frequency-dependent white noise to the model at the same level as measured from the data portraits and finally dispersed the fake data with a DM of 1778~\dmunit.  Visual inspection verified that the fake data were faithfully rendered.  We used the same method to measure the DM (and phase), which is described in detail in \citet{PDR14}.  In summary, the measured DMs were always in accord and unbiased, and the uncertainties were accurately estimated.  We conclude that the measurement method produces accurate DMs, independent of the model parameters, provided that the model for the data is accurate. 

We assume in our measurements that the phase offsets ($\Delta\phi$) incurred by finite-frequency signals due to propagation through the ionized interstellar medium scale as predicted by the usual cold-plasma dispersion law such that $\Delta\phi \propto \frac{\textrm{DM}}{P_s}\ \nu^{-2}$.  This is certainly the case to first-order even over large, low frequency bandwidths \citep{Hassall12}.  However, to the extent that we understand the ISM to be \textit{in}homogeneous --- after all, we do observe pulse broadening --- then it is anticipated that the simple $\nu^{-2}$ dependence will be an insufficient description at some level for broadband DM measurements.  When an inhomogeneous medium causes multi-path propagation of radio waves where the path depends on frequency, the sampled column density of free electrons (the DM) will also be a function of frequency.  Thus, we are left with the intriguing possible explanation that the DM inconsistencies we are seeing are the consequence of imposing a $\nu^{-2}$ dispersion law onto a frequency-dependent DM (DM($\nu$)) due to an inhomogeneous ISM\footnote{This is opposed to other supposed origins of DM($\nu$) relating to magnetospheric propagation effects or magnetic sweepback, which would likely have different statistics from an ISM induced DM($\nu$); see \citet{Hassall12} or \citet{Ahuja07} for an overview.}.

To our knowledge, the most recent claim for having observed frequency-dependent DMs was reported in \citet{Ahuja07} for the slow, low DM pulsars B0329$+$54 and B1642$-$03, although they observed lower DMs at lower frequencies.  However, the authors only made one set of simultaneous pairs of dual-frequency measurements per pulsar.  We argue that to confidently segregate the effects of profile evolution, DM variations with time (DM($t$)), DM($\nu$), and other potential confounding factors, many epochs of simultaneous, wideband (large fractional bandwidth) observations of a stable, preferably high DM pulsar need to be made.  A similar recommendation was recently made by \citet{Cordes15} in their detailed study of frequency-dependent DMs, which makes theoretical predictions for the characteristic timescales and sizes of DM($\nu$) effects.

In their treatment of the problem, \citet{Cordes15} predict the minimum scale of DM variations about a mean value,
\begin{equation}
\label{cordes_5}
\overline{\textrm{DM}}_{\textrm{rms}} \sim \phi_F^2/\lambda r_e \sim 3.84 \times 10^{-8}~\textrm{cm}^{-3}~\textrm{pc}~\nu_{\textrm{GHz}}~\phi_F^2,
\end{equation}
where $\nu_{\textrm{GHz}}$ is the frequency in GHz and $\phi_F$ is the size of the phase perturbations over the Fresnel scale, $l_F = \sqrt{(cD)/(2\pi\nu)}$, for the speed of light $c$ and source distance $D$.  For \gcm, which is in the strong scattering regime, $\phi_F$ will be very large.  We estimate it from their prescription,
\begin{equation}
\label{cordes_14}
\phi_F(\nu) \approx 9.6~\textrm{rad}~\Big(\frac{\nu/\Delta\nu_d}{100}\Big)^{5/12},
\end{equation}
where $\Delta\nu_d$ is the scintillation bandwidth, which is readily estimated from our scattering measurements as $\sim$ 1.16/(2$\pi\tau(\nu)$).  For 1.4 and 2.4~GHz, we find $\overline{\textrm{DM}}_{\textrm{rms}} \sim~$10 and 5~\dmunit, respectively.  These can be compared to the RMS DM values as measured in L- and S-band of $\sim$~9 and 7~\dmunit, respectively.  The characteristic spatial size for the DM differences near 2~GHz will be several Fresnel scales, which can be converted to a characteristic time by using the recently measured proper motion of 236~km~s$^{-1}$ \citep{Bower15}.  For our range of frequencies, the characteristic timescale associated with the Fresnel scale size is $\sim$3~hr, comparable to the separation between the observations on a given epoch.  Therefore, it may be that the small temporal gap between the observations contributes somewhat to the difference in the DMs, but we certainly do expect that the DMs vary significantly on different days, separated by many Fresnel timescales.

Finally, \citet{Cordes15} make a prediction for the observed RMS difference between DMs at frequencies $\nu$ and $\nu'$,
\begin{equation}
\label{cordes_11}
\sigma_{\overline{\textrm{DM}}}(\nu,\nu') \approx 4.42 \times 10^{-5}~\textrm{cm}^{-3}~\textrm{pc}~F_\beta(r)~\Big(\frac{\nu\phi_F^2}{1000}\Big),
\end{equation}
where we have ignored a geometric factor of order unity and the function $F_\beta$ contains the frequency dependence for $r \equiv \nu/\nu'$, given the power-law index $\beta$ for the wavenumber spectrum of density inhomogeneities.  For $\nu = 2.4$~GHz and $\nu' = 1.4$~GHz, $\sigma_{\overline{\textrm{DM}}} \sim 4$~\dmunit, compared to our observed RMS difference of $\sim$6~\dmunit.

That the predicted and observed values are similar may be coincidence, but we note the corroborating facts that \gcm\ is the highest DM pulsar, is relatively bright, highly scattered, has a simple, easily modeled profile, and does not show significant profile evolution or stochastic profile variability (at least in these observations).  Furthermore, we verified that our measurement method produces inconsistent (and biased) DMs between the bands by introducing non-$\nu^{-2}$ phase delays into our fake data simulations described earlier.  After ruling out the other potential sources for the inconsistent DMs, we suggest that \gcm\ may have an observable frequency-dependent dispersion measure.

A potential counter argument is that over many Fresnel timescales, one expects the sign of the DM differences to change, such that the observed low frequency DM becomes smaller than the high frequency DM.  Between the small number and low density of epochs, the potentially incorrect portrait model, and ISM uncertainties (the predictions here are based on a thin-screen model with a Kolmogorov spectrum of density perturbations, which is partly supported by the findings in \citet{Bower14}), it is conceivable that this observation is not inconsistent with a frequency-dependent DM as described.

Determining whether or not a difference in DM as seen in two frequency bands is intrinsically a DM($\nu$) effect is complicated by the issues described above, and with only five measurements we obviously cannot draw any definite or statistical conclusions, but future studies could potentially disentangle the evolution of DM($t$,$\nu$), the profile, and other ISM parameters.  One strategy, as \citet{Cordes15} note, is to model the frequency dependence of the dispersive delays as something other than $\nu^{-2}$.  This should be done for many epochs, at least as long as the timescale for refractive scintillation, over which time the specific frequency dependence of the average DM remains stable.  For \gcm, this timescale is potentially many years. 


\section{Summary \& Discussion}
\label{sum_disc}

In this paper we have presented multi-epoch, multi-frequency wideband GBT observations of the Galactic Center radio magnetar \gcm\ at 1.5, 2.0, and 8.9~GHz from the first $\sim$100~days after it was discovered.  After its initial X-ray burst on 25 April 2013, \gcm\ underwent two additional bursts in the course of our observations.  For two epochs, during which time we collected data from three radio bands, we also have simultaneous X-ray observations taken with \cxo.  An analysis of the radio data, as well as a joint analysis with the X-ray data, yielded a few noteworthy results.

\begin{enumerate}
\item We found no anomalous radio bursts or giant-pulse-like individual pulses in any of the X-band observations.  Similarly, the smooth transitioning of the X-band profile between three broad categories seems to have also been unperturbed by the \swift-detected X-ray bursts.

\item Our simple radio timing analysis corroborates the findings of \citet{Kaspi14}, which are also supported by \citet{Zelati15}.  We presented the absolute alignment of the three radio and 0.3-8~keV profiles.  The near-alignment of the radio components with the X-ray profile is similar to the two other radio magnetars that have published alignments.  We also make note of a possible transient X-ray feature from \citet{Zelati15} because of its proximity to the phase of radio emission located $\sim0.2$ in phase preceding the peak in the X-ray profile.

\item The evolution of our early radio flux measurements, showing a relatively stable growth from around the time of the initial outburst, is consistent with the continued GBT X-band observations presented in \citet{Lynch14} and with what they have called a ``stable~state''\footnote{While our observations were taken over a shorter range of time (about a third), our cadence of observations is comparable to theirs taken during the ``erratic~state'', the onset of which was apparently unrelated to X-ray bursts.}.  The combination of the gradual flux evolution with the simple timing and profile variability results leads us to extrapolate \gcm's ``stable~state'' back to the time of its initial burst.

\item The shape of \gcm's low frequency radio spectrum is potentially positive or flat, whereas it shows a ``typical'' spectral index of $\sim -1.4$ between $\sim$2 and 9~GHz, at least during a brief period $\sim$100~days after its initial outburst, around the times of two later X-ray bursts.  This steep spectral index might indicate a different magnetospheric configuration during these times, although the evolving spectra may be a result of environmental factors and free-free absorption \citep{Lewandowski15b}.  The possible variability of $\gamma$ means that dedicated observations covering several higher frequency bands need to be carried out over many epochs to confirm this \citep[cf.][]{Torne15}.

\item We made wideband models of \gcm's low frequency radio ``portrait'' to measure the scattering timescale, scattering index, and the DM as a function of time.  Our average measurements are consistent with what has been published in \citet{Spitler14}, though the ISM parameters may be variable.  Time-variable scattering parameters would complicate the predicted sensitivities of future pulsar surveys of the Galactic Center.  Lastly, we make a suggestion that our discrepant, nearly simultaneously determined DMs are a manifestation of an ISM-induced frequency-dependent dispersion measure, and that future observations could make a case study out of \gcm\ to investigate DM($\nu$) --- provided the pulsar remains visible and stable.
\end{enumerate}

\gcm\ shares several (but not all) properties with the other three radio magnetars, J1809$-$1943, J1550$-$5418, and J1622$-$4950 \citep{Camilo06,Camilo07c,Levin10}.  Common properties of the pulsed radio emission from magnetars are: a) a delay in the appearance of the radio emission after the X-ray outburst onset, b) variable pulse profiles and radio flux on timescales from hours to days, c) a large rotational (spin-down) luminosity with respect to the quiescent X-ray luminosity, d) a decrease of the radio flux as the X-ray flux decays, and e) a flat radio spectrum over a wide range of frequencies.  \gcm\ grossly shares the first three properties with the rest of its class.  However, while in all other cases the radio flux was observed to decay as the X-ray outburst was fading, the long-term radio and X-ray flux evolution of \gcm\ is at variance with this trend.  The radio flux shows a re-brightening hundreds of days after the outburst onset and the X-ray emission is decaying very slowly, challenging current crustal cooling models \citet{Zelati15}.  Furthermore, the recently published flux measurements by \citet{Torne15} taken one year after those presented here suggest that the 8.35~GHz flux remained stable at the $\sim$3~mJy level over thirty days.  Another interesting peculiarity of this radio magnetar was the steep (and possibly free-free absorbed) radio spectrum seen in our observations, though the more recent observations in \citet{Torne15} suggest that the spectrum has since flattened.

Of particular interest is \gcm's low quiescent luminosity compared to its high rotational power \citep[$L_{\textrm{X,qui}}/L_{\textrm{rot}} < 1$;][]{Rea13}.  This peculiarity of the four radio magnetars, which is at variance with canonical magnetars (for which the fact that $L_{\textrm{X,qui}}/L_{\textrm{rot}} > 1$ has always been used as proof of their magnetically dominated emission \citep{MereghettiStella95,TD95,Mereghetti08}), has been viewed as evidence for a similar mechanism powering the radio emission from magnetars and normal pulsars alike.  In fact, while normal radio pulsars have primarily dipolar-dominated magnetic fields ($B_\textrm{p}$), magnetars have a substantial toroidal component ($B_\phi$) that is present in both the internal and external fields.  This toroidal component is the main reason for their quiescent X-ray luminosities, hot surface temperatures, flaring emission, and outburst activity \citep{Thompson02,Beloborodov09}.  For a fixed dipolar field, the internal toroidal field has no significant effect on the luminosity unless $B_\phi > B_\textrm{p}$, as is the case for most magnetars \citep{Vigano13}.  Both radio magnetars and high-$B$ radio pulsars have systematically lower toroidal fields and higher rotational energies than typical magnetars; this is in agreement with the former being fainter in quiescence and having a softer X-ray spectrum \citep[a lower crustal toroidal field results in less heating produced by Joule dissipation in the crust,][]{Pons09}.  As for the energy powering the radio emission, simulations of high dipolar field pulsars that have a small toroidal component showed that the particle acceleration and subsequent ignition of the cascade process could proceed as it does in normal pulsars, successfully reaching the open-field line region and generating pulsed radio emission \citep{MedinLai10}.  On the other hand, for an extremely strong toroidal component, it is expected that the particle cascades cannot reach the open-field lines due to the powerful currents formed as a consequence of the twisted magnetosphere.  Radio magnetars might lie in between, having a high enough rotational energy to power pair cascades as in normal pulsars, but also having toroidal components lower than typical magnetars, resulting in lower quiescent X-ray luminosities.

In the above picture, the possible radio flux increase, the steep spectrum, and the slow cooling of the X-ray outburst might be explained by the presence of a strongly twisted bundle, which can account for the radio emission and the additional heating by particles slamming onto the surface.  If the radio emission is generated by acceleration of particles only in this part of the magnetosphere, then the radio flux and the X-ray flux might be unrelated.  In particular, untwisting of the bundle during the outburst decay might induce fewer currents blocking the pair cascade generation, hence more radio emission from this region.  However, these are only speculative, plausible hypotheses.  Proof of this scenario would need a longer monitoring of the radio and X-ray emission, as well as detailed magnetohydrodynamical simulations of particle acceleration and pair cascades in a strongly magnetized and twisted bundle.

\acknowledgments 
The authors would like to thank S. Ransom and P. Demorest for useful discussions and comments, as well as A. Bilous for significant contributions and a review of the manuscript.  TTP would also like to acknowledge the hospitality of the astronomers at the Osservatorio Astronomico di Cagliari, where this paper began.  TTP is supported in part by a National Science Foundation PIRE Grant (0968296) through NANOGrav and is a graduate student at the National Radio Astronomy Observatory, a facility of the National Science Foundation operated under cooperative agreement by Associated Universities, Inc.  PE acknowledges a Fulbright Research Scholar grant administered by the U.S.--Italy Fulbright Commission and is grateful to the Harvard--Smithsonian Center for Astrophysics for hosting him during his Fulbright exchange.  The Fulbright Scholar Program is sponsored by the U.S. Department of State and administered by CIES, a division of IIE.  DH acknowledges support from \cxo\ X-ray Observatory Award Number GO3-14121X, operated by the Smithsonian Astrophysical Observatory for and on behalf of NASA under contract NAS8-03060, and also by NASA \swift\ grant NNX14AC30G.  NR and FCZ are supported by an NWO Vidi Grant (PI: N. Rea) and by the European COST Action MP1304 (NewCOMPSTAR).  NR acknowledges support by grants AYA2012-39303 and SGR2014-1073. 

{\it Facilities:} \facility{CXO (ACIS)}, \facility{GBT (GUPPI)}

\bibliography{apj-jour,GC_magnetar}

\end{document}